\documentclass[prl,aps,twocolumn,showpacs]{revtex4}
\usepackage{mathrsfs}
\usepackage{amsfonts}
\usepackage{txfonts}
\usepackage{amssymb}
\usepackage{graphicx}
\newcommand{\ket}[1]{|#1\rangle}
\newcommand{\bra}[1]{\langle #1|}

\begin{document}
\title{Non-Adiabatic Holonomic Quantum Computation in Decoherence-Free Subspaces}
\author{G. F. Xu$^1$, J. Zhang$^1$, D. M. Tong$^{1}$\footnote{Email: tdm@sdu.edu.cn},
Erik Sj\"{o}qvist$^{2,3}$, L. C. Kwek$^3$}
\affiliation{$^1$Department of Physics, Shandong University, Jinan
250100, China \\
$^2$Department of Quantum Chemistry, Uppsala University, Box 518,
Se-751 20 Uppsala, Sweden \\
$^3$Center for Quantum Technologies, National
University of Singapore, Science Drive 2, Singapore 117543}
\date{\today}
\begin{abstract}
Quantum computation that combines the coherence stabilization virtues of
decoherence-free subspaces and the fault tolerance of geometric holonomic
control is of great practical importance. Some schemes of adiabatic holonomic
quantum computation in decoherence-free subspaces have been proposed in
the past few years. However, non-adiabatic holonomic quantum computation in
decoherence-free subspaces, which avoids long run-time requirement
but with all the robust advantages, remains an open problem. Here, we
demonstrate how to realize non-adiabatic
holonomic quantum computation in decoherence-free subspaces. By
using only three neighboring physical qubits undergoing collective dephasing
to encode one logical qubit, we realize a universal set of quantum gates.
\end{abstract}
\maketitle
\date{\today}

The discovery of geometric phase \cite{Berry,Aharonov} and quantum holonomy
\cite{Wilczek,Anandan} accompanying evolutions of quantum systems has unveiled
important geometric structures in the description of physical states.
These structures show that the twisting of subspaces, e.g., eigenspaces of adiabatically varying Hamiltonian,
can be used to manipulate quantum states in a robust manner. This is the initial idea of holonomic
quantum computation (HQC), first proposed by Zanardi and Rasetti \cite{zanardi99}. HQC
has emerged as a key tool to implement quantum gates acting on sets of quantum
bits (qubits). As is well known, errors in the control process of a quantum system
are one main practical difficulty in building a quantum computer, and propagation of these errors may quickly spoil the whole quantum computational process. Since HQC is fault tolerant with respect to certain types of errors in the control process, it has been used to realize
robust quantum computation \cite{Sjoqvist,Golovach,Wu,Feng1,Duan,Jones,WangXB,Zhu,Oreshkov1,Thomas,Cen,Zhang}.

Besides errors produced in the control process, decoherence is another main practical difficulty
in building a quantum computer. Decoherence is caused by the inevitable interaction between
the computational system and its environment. It collapses the desired coherence of the system
and may thereby be detrimental to the efficiency of quantum computation. Protecting
qubits from the effects of decoherence is a vital requirement for any quantum
computer implementation. Various strategies have been proposed to protect quantum information against decoherence.
Among the them, decoherence-free subspaces (DFSs) provide a promising way to
avoid quantum decoherence \cite{lidar98}. The basic idea of DFSs is to utilize the symmetry structure
of the interaction between the system and its environment. Information is encoded in a subspace
of the Hilbert space of a system, over which the dynamics is unitary. DFSs have been experimentally
realized in many physical systems \cite{Kielpinski,Kwiat,Mohseni,Ollerenshaw,Bourennane}.

To protect quantum information from both errors produced in the control process and
decoherence caused by the environment, quantum gates that combine the coherence
stabilization virtues of DFSs and the fault tolerance of geometric holonomic control are of
great practical importance. To this end, schemes of HQC in DFSs have been proposed recently
\cite{Wu,Cen,Zhang}. Wu $et.~al$ \cite{Wu} proposed the first scheme of adiabatic HQC in
DFSs, in which one logical qubit is encoded by four neighboring physical qubits and
the quantum holonomies are accumulated by adiabatically changing the couplings between
the qubits along dark states. The scheme is robust against collective dephasing and some
stochastic errors. Yet, the requirement of adiabatic control of four neighboring physical qubits
undergoing collective dephasing is an experimental challenge. All other schemes
that can realize a universal set of holonomy quantum gates in DFSs are based on
adiabatic evolution too, and they met the same problem of long-run time requirement.

In this Letter, we develop a scheme for non-adiabatic universal holonomic quantum computation in
decoherence-free subspaces. Our proposal avoids the long run-time requirement but shares
all the robust advantages of its adiabatic counterpart. An additional attractive feature
of this non-adiabatic setting is that only three neighboring physical qubits undergoing
collective dephasing are needed to encode one logical qubit. We further demonstrate that
three neighboring physical qubits is the minimal number for realizing non-adiabatic HQC
in DFSs, although two neighboring physical qubits may construct the minimal DFS.

Before proceeding further, we explain how quantum holonomy may arise in non-adiabatic
unitary evolution. Consider a quantum system described by an $N-$dimensional state
space and exposed to the Hamiltonian $H(t)$. Assume there is a time-dependent $L-$dimensional
subspace $\mathcal{S} (t)$ spanned by the orthonormal basis vectors $\{ \ket{\phi_k(t)} \}_{k=1}^L$
at each instant $t$. Here, $\ket{\phi_k(t)}$ satisfy the Schr\"odinger equation
$i\ket{\dot\phi_k(t)}=H(t)\ket{\phi_k(t)}$. That is, $\ket{\phi_k (0)} \rightarrow
\ket{\phi_k(t)} = U(t,0)\ket{\phi_k(0)}$ with the time evolution operator $U(t,0) =
{\bf T} \exp{\left(-i\int_0^tH(t')dt'\right)}$, ${\bf T}$ being time ordering. One may conclude that the unitary transformation
$U(\tau,0)$ is a holonomy matrix acting on the $L-$dimensional subspace $\mathcal{S}(0)$
spanned by $\{ \ket{\phi_k(0)} \}_{k=1}^L$ if $\ket{\phi_k (t)}$ satisfy the following
requirements:
\begin{eqnarray}
& \textrm{(i)} & \ \ \sum_{k=1}^L\ket{\phi_k (\tau)} \bra{\phi_k (\tau)} =
\sum_{k=1}^L \ket{\phi_k (0)} \bra{\phi_k (0)} , \\
& \textrm{(ii)} & \ \ \bra{\phi_k (t)}H(t)\ket{\phi_l (t)}=0, \ k,l = 1, \ldots ,L.
\label{conditions}
\end{eqnarray}
To verify that $U(\tau,0)$ is a holonomy matrix acting on $\mathcal{S} (0)$, we first note
that condition (i) entails that the subspace undergoes cyclic evolution, i.e., we can introduce
a set of the auxiliary bases $\ket{\nu_k (t)}$ of $\mathcal{S}(t)$ with the property
\begin{eqnarray}
\ket{\nu_k (\tau) }= \ket{\nu_k (0)} = \ket{\phi_k (0)}, \ k=1, \ldots ,L.
\label{nu}
\end{eqnarray}
Note that $\ket{\nu_k (t)}$ need not satisfy the Schr\"odinger equation, and therefore such bases
can always be found \cite{nuit}. By the aid of $\ket{\nu_k (t)}$, $\ket{\phi_k(t)}$
may be expressed as
\begin{eqnarray}
\ket{\phi_k (t)}=\sum_{l=1}^L \ket{\nu_l (t)} C_{lk} (t) ,
\label{epan}
\end{eqnarray}
where $C_{kl} (t)$ are time dependent coefficients. Substituting Eq. (\ref{epan}) into the
Schr\"odinger equation yields
\begin{eqnarray}
\frac{d}{dt}C_{lk}(t) = i \sum_{m=1}^L \big( A_{lm} (t) - K_{lm} (t) \big) C_{mk} (t),
\label{sch2}
\end{eqnarray}
where $A_{kl} (t) = i \bra{\nu_k (t)} \frac{d}{dt} \ket{\nu_l (t)}$, and $K_{kl}(t) =
\bra{\nu_k (t)} H(t)\ket{\nu_l (t)}$. Condition (ii) is equivalent to $K_{kl} (t) = 0$, i.e.,
the Hamiltonian vanishes on $\mathcal{S} (t)$ and hence $C(t)={\bf T}\exp{\left(i\int_0^tA(t')dt'\right)}$.
The matrix $A(t)$ transforms as a proper gauge potential under the change
$\ket{\nu_k (t)} \rightarrow \sum_{l=1}^L \ket{\nu_l} V_{lk} (t)$, where $V(t)$ is any
unitary once differentiable $L \times L$ matrix such that $V(\tau) = V(0)$.
At time $t=\tau$, there is $C(\tau)={\bf P} e^{i\oint \mathcal{A}}$, where $\mathcal{A} = Adt$ is the connection one-form and ${\bf P}$ is path ordering. From Eq. (\ref{epan}), we have $\ket{\phi_k(\tau)}=\sum_{l=1}^L\ket{\nu_l(\tau)}C_{lk}(\tau)=\sum_{l=1}^L\ket{\phi_l(0)}C_{lk}(\tau)$. It indicates that $C(\tau)$ is just the transformation matrix from initial states to final states in the subspace considered. Hence, we finally obtain
\begin{eqnarray}
U(\tau)\equiv C(\tau)={\bf P} e^{i\oint \mathcal{A}}.\label{utc}
\end{eqnarray}
Equation (\ref{utc}) shows that $U(\tau)$ is a holonomy matrix in the space spanned by $\{\ket{\phi_k(t)}\}_{k=1}^L$.

Let us now elucidate our physical model. The computational system consists of $N$ physical
qubits interacting collectively with a dephasing environment. The Hamiltonian of the system
reads
\begin{eqnarray}
H=\sum_{k<l}(J_{kl}^xR_{kl}^x+J_{kl}^yR_{kl}^y),\label{h}
\end{eqnarray}
where $J_{kl}^x$ and $J_{kl}^y$ are controllable coupling constants, which are driven to
enact the quantum computation, and
\begin{eqnarray}
R_{kl}^x=\frac{1}{2}(\sigma_k^x\sigma_l^x+\sigma_k^y\sigma_l^y), \ \
R_{kl}^y=\frac{1}{2}(\sigma_k^x\sigma_l^y-\sigma_k^y\sigma_l^x) .
\end{eqnarray}
The operators $R_{kl}^x$ and $R_{kl}^y$ are XY and Dzialoshinski-Moriya \cite{Dzialoshinski,Moriya}
interaction terms, where $\sigma_k^x$ ($\sigma_k^y$) represents the Pauli $X$ ($Y$) operator
acting on the $k$th qubit.  A variety of quantum
systems, including trapped ions and quantum dots, can be described by this Hamiltonian
\cite{Mozyrsky,Imamoglu,Zheng,Sorensen}. The major source of
decoherence in the quantum system is dephasing. The effect of the dephasing environment on
the $N-$qubit system is described by the interaction Hamiltonian,
\begin{eqnarray}
H_I=\big(\sum_k\sigma_k^z\big)\otimes{B},
\end{eqnarray}
where $\sigma_k^z$ is the Pauli $Z$ operator acting on the $k$th qubit, and $B$ is an arbitrary
environment operator. The symmetry of the interaction implies that there exists a DFS that can
be used to protect quantum information against decoherence. Our aim is to find a realization
of non-adiabatic HQC in this DFS.

We begin by showing that two physical qubits are not sufficient to realize decoherence-free
non-adiabatic HQC in the presence of a dephasing environment. For a two-qubit system,
the corresponding DFS is spanned by $\big\{\ket{01}, \ket{10}\big\}$.  In order to protect
the quantum gates from decoherence, logical qubits must be encoded in this DFS, and the
state of the system must be kept within the subspace during the whole evolution. Thus, the
DFS itself must be an invariant subspace during the system's evolution. In addition, to ensure
that the gates are holonomic, condition (ii) must be satisfied, i.e., $\bra{k} U(t,0)^{\dagger}
H(t) U(t,0) \ket{l} = 0$ for $k,l = 01, 10$. This is equivalent to $\bra{k} H (t)\ket{l }= 0$ since
the DFS is an invariant subspace. Thus, $H(t)=0$ in the subspace and it follows that one
cannot realize non-adiabatic HQC in the DFS of two physical qubits since there is no nontrivial
Hamiltonian to meet conditions (i) and (ii) above.

For three physical qubits interacting collectively with the dephasing environment, there
exists a three-dimensional DFS
\begin{eqnarray}
\mathcal{S}^D = \text{Span} \big\{ \ket{100}, \ket{010}, \ket{001} \big\}.
\end{eqnarray}
We encode a logical qubit in the subspace
\begin{eqnarray}
\mathcal{S}^L = \text{Span} \big\{ \ket{010}, \ket{001} \big\},
\end{eqnarray}
and denote the computational basis elements as $\ket{0}_L = \ket{010}, \ket{1}_L = \ket{001}$.
Clearly, $\mathcal S^L$ is a subspace of $\mathcal S^D$ and the remaining vector $\ket{100}$
is used as ancillae, denoted as $\ket{a}=\ket{100}$ for convenience. In the following paragraphs,
we utilize the DFS of three physical qubits to implement non-adiabatic HQC. To this end, we
need to generate two noncommuting single-qubit gates and one nontrivial two-qubit gate.

Firstly, we demonstrate how to realize the one-qubit holonomic gate
\begin{eqnarray}
U_{xz}(\phi_1) = X_Le^{i\phi_1Z_L} .
\end{eqnarray}
Here, $X_L=\ket{0}_L\bra{1}_L+\ket{1}_L\bra{0}_L$, $Z_L=\ket{0}_L\bra{0}_L-\ket{1}_L\bra{1}_L$
are the Pauli operators of the logical qubit and $\phi_1$ is an arbitrary phase. In the computational
basis $\{\ket{0}_L,\ket{1}_L\}$, the gate reads
\begin{eqnarray}
U_{xz}(\phi_1)=\left(\begin{array}{cc} 0& e^{-i\phi_1}\\ e^{i\phi_1}&0\end{array}\right).\label{uxz}
\end{eqnarray}
In order to realize $U_{xz}(\phi_1)$, we  set
$J_{12}^x=J_1\cos\frac{\phi_1}{2}$,
$J_{12}^y=-J_1\sin\frac{\phi_1}{2}$,
$J_{13}^x=-J_1\cos\frac{\phi_1}{2}$,
$J_{13}^y=-J_1\sin\frac{\phi_1}{2}$,
and all other $J^{x(y)}_{kl}$ to zero, where $J_1$ is a time-independent parameter \cite{j123}. The Hamiltonian then reads
\begin{eqnarray}
H_1 = J_1 \left[ (R_{12}^x - R_{13}^x) \cos\frac{\phi_1}{2} -
(R_{12}^y+R_{13}^y) \sin\frac{\phi_1}{2} \right].
\label{h1}
\end{eqnarray}
$\mathcal{S}^D$ itself is an
invariant subspace of the evolution operator $U_1(t)=e^{-iH_1t}$. In
the basis $\{\ket{a},\ket{0}_L,\ket{1}_L\}$, we have
\begin{eqnarray}
H_1= J_1\left(\begin{array}{ccc}
0 & e^{i\frac{\phi_1}{2}} & -e^{-i\frac{\phi_1}{2}} \\
e^{-i\frac{\phi_1}{2}} & 0 & 0 \\
-e^{i\frac{\phi_1}{2}} & 0 & 0
\end{array}\right)\label{h1matrix} .
\end{eqnarray}
With the expression of $H_1$, we can work out the operator
$U_1(t)$. By choosing the evolution time $\tau_1$
such that
\begin{eqnarray}
J_1\tau_1=\frac{\pi}{\sqrt{2}} ,
\label{j1}
\end{eqnarray}
the resulting unitary operator reads
\begin{eqnarray}
U_1(\tau_1)=
\left(\begin{array}{ccc}
-1 & 0 & 0 \\
0 & 0 & e^{-i\phi_1} \\
0 & e^{i\phi_1} & 0
\end{array}\right) .
\end{eqnarray}
Thus, the action of the evolution operator $U_1(\tau_1)$ on the states in the logic subspace
$\mathcal S^L$ is equivalent to that of the transformation $U_{xz}(\phi_1)$.

In order to ensure that the action of $U_1 (\tau_1)$ on $\mathcal{S}^L$ is purely holonomic,
we need to check conditions (i) and (ii). Condition (i) is satisfied since the subspace spanned
by $\{U_1(\tau_1)\ket{0}_L,U_1(\tau_1) \ket{1}_L\}$ coincides with $\mathcal{S}^L$.
Furthermore, as $H_1$ and $U_1(t)$ commute with each other, condition (ii) reduces to
$\bra{k}_L{H_1}\ket{k^{\prime}}_L=0$, where $k,k^{\prime}=0,1$. Thus, both conditions
(i) and  (ii) are satisfied, and $U_1(\tau_1)$ is therefore a one-qubit holonomic gate in the
subspace $\mathcal S^L$, $\mathcal S^L\subset\mathcal S^D$.

Secondly, we demonstrate how to realize the one-qubit holonomic gate
\begin{eqnarray}
U_{zx}(\phi_2)=Z_Le^{i\phi_2X_L},
\end{eqnarray}
where $\phi_2$ is an arbitrary phase. In the computational basis $\{\ket{0}_L,~\ket{1}_L\}$,
we have
\begin{eqnarray}
U_{zx}(\phi_2) = \left( \begin{array}{cc}
\cos\phi_2 & i\sin\phi_2 \\
-i\sin\phi_2 & - \cos\phi_2
\end{array}\right).
\label{uzx}
\end{eqnarray}
To realize $U_{zx}$, we set $J_{12}^y=J_2\sin\frac{\phi_2}{2}$,
$J_{13}^x=-J_2\cos\frac{\phi_2}{2}$, and all other $J^{x(y)}_{kl}$ to zero, where $J_2$ is a time-independent parameter \cite{j123}. The Hamiltonian then reads
\begin{eqnarray}
H_2 = J_2 \left( R_{12}^y\sin\frac{\phi_2}{2} - R_{13}^x \cos\frac{\phi_2}{2} \right).
\label{h2}
\end{eqnarray}
Again $\mathcal {S}_D$ is an invariant subspace of
$U_2(t)=e^{-iH_2t}$. Expressed in the
$\{\ket{a},\ket{0}_L,\ket{1}_L\}$, the resulting time evolution
operator takes the form
\begin{eqnarray}
U_2(\tau_2)=
\left(\begin{array}{ccc}-1& 0& 0\\
0 & \cos\phi_2& i\sin\phi_2\\
0 & -i\sin\phi_2& -\cos\phi_2
\end{array}\right)
\label{ust}
\end{eqnarray}
by choosing the evolution time $\tau_2$ such that
\begin{eqnarray}
J_2\tau_2=\pi.
\label{j2}
\end{eqnarray}
Equation (\ref{ust}) shows that the action of the evolution operator $U_2(\tau_2)$ on
$\mathcal{S}^L$ is equivalent to that of $U_{zx}(\phi_2)$. Its holonomic nature is
demonstrated as above. Thus, $U_2(\tau_2)$ acts as a one-qubit holonomic gate in
the subspace $\mathcal S^L$.

We note that any single-qubit operation can be written as a combination of the following
two types of rotations
\begin{eqnarray}
R_z(\theta)=e^{-i\frac{\theta}{2}\sigma^z}, \ \ R_x(\varphi)=e^{-i\frac{\varphi}{2} \sigma^x},
\end{eqnarray}
where $\theta,\varphi$ are rotation angles and $\sigma^z,\sigma^x$ are Pauli operators.
Equations (\ref{uxz}) and (\ref{uzx}) imply
\begin{eqnarray}
U_{xz}(0)U_{xz}(-\theta/2) = e^{-i\frac{\theta}{2}Z_L}, \ \
U_{zx}(0)U_{zx}(-\varphi/2) =e^{-i\frac{\varphi}{2}X_L},
\end{eqnarray}
where $X_L$ and $Z_L$ are just the Pauli $Z$ and Pauli $X$ operators of the logical qubit.
This proves that $U_{xz}(\phi_1)$ and $U_{zx}(\phi_2)$ can realize any single-qubit rotation.

Thirdly, we demonstrate how to realize a nontrivial two-qubit gate. It is worth noting that the
Hamiltonian in Eq. (\ref{h}) serves single-qubit gates but cannot directly be applied to
implement two-qubit gates. To implement a holonomic two-qubit gate, four-qubit
interactions are needed. Here, we generate the CNOT gate
by means of the Hamiltonian,
\begin{eqnarray}
H_3 = J_3 \left( R_{13}^x R_{45}^x - R_{13}^x R_{46}^x \right) ,
\label{h3}
\end{eqnarray}
where $J_3$ is a time-independent parameter \cite{j123}. The
Hamiltonian $H_3$ is obtained  by setting $J^{xx}_{13,45}=-J^{xx}_{13,46}=J_3$ and all other controllable
four-qubit coupling constants to zero. The choice of $J_3$ is related to the evolution time
$\tau_3$. The requirement for $J_3$ or $\tau_3$ is
\begin{eqnarray}
J_3\tau_3=\frac{\pi}{\sqrt{2}}.\label{j3}
\end{eqnarray}
In this case, $\mathcal S^D\otimes \mathcal S^D$ is a
decoherence-free subspace, in which the small subspace spanned
by $\{\ket{a}\otimes\ket{a},~\ket{0}_L\otimes\ket{0}_L,~\ket{0}_L\otimes\ket{1}_L,
~\ket{1}_L\otimes\ket{0}_L,~\ket{1}_L\otimes\ket{1}_L\}$ is an
invariant subspace of the Hamiltonian $H_3$. In the invariant
subspace, the evolution operator at time $t=\tau_3$ reads
\begin{eqnarray}
U_3(\tau_3)=
\left(\begin{array}{ccccc}-1& 0& 0& 0& 0\\
  0& 1& 0& 0& 0\\
  0& 0& 1& 0& 0\\
  0& 0& 0& 0& 1\\
  0& 0& 0& 1& 0
\end{array}\right).
\end{eqnarray}
Then, the CNOT gate is realized in the subspace $\mathcal{S}^L \otimes \mathcal{S}^L$, i.e., $\text{span}\{\ket{0}_L\otimes\ket{0}_L,\ket{0}_L\otimes\ket{1}_L,
\ket{1}_L\otimes\ket{0}_L,\ket{1}_L\otimes\ket{1}_L\}$.
One may verify that conditions (i) and (ii) are fulfilled too. $U_3(\tau_3)$ plays a
two-qubit holonomic CNOT gate in the subspace $\mathcal{S}^L \otimes \mathcal{S}^L$.

We have succeeded to construct two non-commuting holonomic single-qubit gates $U_{xz}$
and $U_{zx}$ and a holonomic CNOT two-qubit gate in DFSs of a system undergoing collective
dephasing. The three gates compose a universal set of non-adiabatic holonomic quantum gates
in DFSs. It is worth noting that the scheme proposed here is suitable for scaling up the logic qubits. The Hamiltonian to realize the gates of the $n-$th logic qubit has the same structure as $H_1$ or $H_2$ but with the exchanging $R_{12}^{x(y)}\rightarrow R_{3n-2,3n-1}^{x(y)}$ and $R_{13}^{x(y)}\rightarrow R_{3n-2,3n}^{x(y)}$, while the Hamiltonian to realize the CNOT gate between the $m-$th and the $n-$th logic qubits has the same structure as $H_3$ but with the exchanging $R_{13}^{x}R_{45}^{x}\rightarrow R_{3m-2,3m}^{x}R_{3n-2,3n-1}^{x}$ and $R_{13}^{x}R_{46}^{x}\rightarrow R_{3m-2,3m}^{x}R_{3n-2,3n}^{x}$.

In summary, we have put forward a scheme for non-adiabatic holonomic quantum computation
in decoherence-free subspaces. By using only three neighboring physical qubits undergoing
collective dephasing to encode one logical qubit, we realize a universal set of quantum gates.
Our scheme combines the coherence stabilization virtues of decoherence-free subspaces and
the fault tolerance of geometric holonomic control. Comparing with the previous schemes, our
scheme has removed the long run-time requirement in the adiabatic evolution and can avoid
the extra errors and decoherence involved due to long time evolution. Since the Hamiltonian in the
scheme may be independent of time, our scheme seems promising experimental implementation,
which may shed light on the applications of holonomic quantum computation in decoherence-free
subspaces.

\section*{Acknowledgments}
This work was supported by NSF China with No.11175105 and the National Basic Research
Program of China (Grant No. 2009CB929400). Tong and Sj\"{o}qvist acknowledge support from the National
Research Foundation and the Ministry of Education (Singapore).


\begin{thebibliography}{99}
\bibitem{Berry} M. V. Berry,
Proc. R. Soc. London A {\bf 392}, 45 (1984).
\bibitem{Aharonov} Y. Aharonov and J. Anandan,
Phys. Rev. Lett. {\bf 58}, 1593 (1987).
\bibitem{Wilczek} F. Wilczek and A. Zee,
Phys. Rev. Lett. {\bf 52}, 2111 (1984).
\bibitem{Anandan} J. Anandan,
Phys. Lett. A {\bf 113}, 171 (1988).
\bibitem{zanardi99} P. Zanardi and M. Rasetti,
Phys. Lett. A {\bf 264}, 94 (1999).
\bibitem{Jones} J. A. Jones, V. Vedral, A. Ekert, and G. Castagnoli,
Nature (London) {\bf 403}, 869 (2000).
\bibitem{Duan} L. M. Duan, J. I. Cirac, and P. Zoller,
Science {\bf 292}, 1695 (2001).
\bibitem{WangXB} X. B. Wang and K. Matsumoto,
Phys. Rev. Lett., {\bf 87}, 097901 (2001).
\bibitem{Zhu} S. L. Zhu and Z. D. Wang, Phys. Rev. Lett. {\bf 91}, 187902 (2003).
\bibitem{Wu} L. A. Wu, P. Zanardi, and D. A. Lidar, Phys. Rev. Lett. {\bf 95}, 130501 (2005).
\bibitem{Cen} L. X. Cen, Z. D. Wang, and S. J. Wang, Phys. Rev. A
{\bf 74}, 032321 (2006).
\bibitem{Zhang} X. D. Zhang, Q. H. Zhang, and Z. D. Wang, Phys. Rev.
A {\bf 74}, 034302 (2006).
\bibitem{Feng1} X. L. Feng, C. F. Wu, H. Sun, and C.H. Oh, Phys. Rev. Lett. {\bf 103}, 200501 (2009).
\bibitem{Oreshkov1} O. Oreshkov, T. A. Brun and D. A. Lidar,
Phys. Rev. Lett. {\bf 102}, 070502 (2009).
\bibitem{Golovach} V. N. Golovach, M. Borhani, and D. Loss,
Phys. Rev. A {\bf 81}, 022315 (2010).
\bibitem{Sjoqvist} E. Sj\"{o}qvist, D. M. Tong, B. Hessmo, M. Johansson, and K. Singh,
arXiv: 1107, 5127, (2011).
\bibitem{Thomas} J. T. Thomas, M. Lababidi, and M. Z. Tian,
Phys. Rev. A {\bf 84}, 042335 (2011).
\bibitem{lidar98} D. A. Lidar, I. L. Chuang, and K. B. Whaley,
Phys. Rev. Lett. {\bf 81}, 2594 (1998).
\bibitem{Kwiat} P. G. Kwiat, A. J. Berglund, J. B. Altepeter, and A. G. White,
Science {\bf 290}, 498 (2000).
\bibitem{Kielpinski} D. Kielpinski, V. Meyer, M. A. Rowe, C. A. Sackett, W. M. Itano, C. Monroe, and
D. J. Wineland,
Science {\bf 291}, 1013 (2001).
\bibitem{Mohseni} M. Mohseni, J. S. Lundeen, K. J. Resch, and A. M. Steinberg,
Phys. Rev. Lett. {\bf 91}, 187903 (2003).
\bibitem{Ollerenshaw}J. E. Ollerenshaw, D. A. Lidar, and L. E. Kay, Phys. Rev. Lett. {\bf 91}, 217904 (2003).
\bibitem{Bourennane}M. Bourennane, M. Eibl, S. Gaertner, C. Kurtsiefer, A. Cabello, and H. Weinfurter, Phys. Rev. Lett. {\bf 92}, 107901 (2004).
\bibitem{nuit}In fact,  $\{\ket{\nu_i(t)}\}$ can be taken as a linear combination of $\{\ket{\phi_i(t)}\}$ with time dependent coefficients. Equation (\ref{nu}) can be fulfilled by properly choosing the coefficients.
\bibitem{Dzialoshinski}L. Dzialoshinski, J. Phys. Chem. Solids {\bf
4}, 241, (1958).
\bibitem{Moriya}T. Moriya, Phys. Rev. Lett. {\bf 4}, 228 (1960).
\bibitem{Mozyrsky}D. Mozyrsky, V. Privman, and M. L. Glasser, Phys. Rev. Lett. {\bf 86}, 5112 (2001).
\bibitem{Imamoglu}A. Imamoglu, D. D. Awschalom, G. Burkard, D. P. DiVincenzo, D. Loss, M. Sherwin, and A. Small, Phys. Rev. Lett. {\bf 83}, 4204 (1999).
\bibitem{Zheng}S. B. Zheng and G. C. Guo, Phys. Rev. Lett. {\bf 85}, 2392 (2000).
\bibitem{Sorensen}A. S{\o}rensen and K. M{\o}lmer, Phys. Rev. A {\bf 62}, 022311 (2000).
\bibitem{j123} $J_i$, $i=1,2,3$ are not necessary to be constants. If $J_i$ are taken as time-dependent parameters, terms $J_i\tau_i$ in Eqs. (\ref{j1}), (\ref{j2}) and (\ref{j3}) need only to be replaced by $\int_0^\tau J_i dt$.
\end{thebibliography}
\end{document}